\documentclass[12pt]{article}
\usepackage{graphicx}

\newcommand\pubnumber{SNSN-323-63}
\newcommand\pubdate{\today}

\def\napoli{Universidad de Oviedo, SPAIN}

\def\Title#1{\begin{center} {\Large #1 } \end{center}}
\def\Author#1{\begin{center}{ \sc #1} \end{center}}
\def\Address#1{\begin{center}{ \it #1} \end{center}}

\newcommand\pubblock{\rightline{\begin{tabular}{l} \pubnumber\\
         \pubdate  \end{tabular}}}
\newenvironment{Abstract}{\begin{quotation}  }{\end{quotation}}
\newenvironment{Presented}{\begin{quotation} \begin{center} 
             PRESENTED AT\end{center}\bigskip 
      \begin{center}\begin{large}}{\end{large}\end{center} \end{quotation}}

\def\beq{\begin{equation}}
\def\eeq#1{\label{#1}\end{equation}}
\def\eeqn{\end{equation}}

\def\beqa{\begin{eqnarray}}
\def\eeqa#1{\label{#1}\end{eqnarray}}
\def\eeqan{\end{eqnarray}}

\let\bar=\overbar

\def\Dslash{\not{\hbox{\kern-4pt $D$}}}
\def\dslash{\not{\hbox{\kern-2pt $\del$}}}

\def\msb{{\bar{\ssstyle M \kern -1pt S}}}

\begin{document}
\begin{titlepage}
\pubblock

\vfill
\Title{Latest results from CMS}
\vfill
\Author{Enrique Palencia Cortezon \\(on behalf of the CMS Collaboration)}
\Address{\napoli}
\vfill
\begin{Abstract}
A summary of the latest results released by the CMS Collaboration during the Summer of 2016 is presented.
\end{Abstract}
\vfill
\begin{Presented}
9th International Workshop on Top Quark Physics\\
Olomouc, Czech Republic, Spetember 19-23, 2016
\end{Presented}
\vfill
\end{titlepage}
\def\thefootnote{\fnsymbol{footnote}}
\setcounter{footnote}{0}

\section{Introduction}

During the summer of 2016, the CMS Collaboration has released a large variety of new results  based on the 13 TeV proton-proton collisions data  recorded by the CMS detector~\cite{CMS} at the CERN LHC untill June 2016. The analysis described below use an integrated luminosity of $\leq$12.9 fb$^{-1}$ and cover a large spectra, from precission measurements in Standard Model (SM) physics to search for new physics scenarios.

\section{Standard Model}

The first measurement of the differential cross sections for a W boson produced in association with jets in proton-proton collisions at a centre-of-mass energy of 13 TeV is presented~\cite{smp16005}. The data used correspond to an integrated luminosity of 2.5 fb$^{-1}$. The differential cross sections are measured using the $\mu$ decay mode of the W boson as a function of the exclusive and the inclusive jet multiplicities up to a multiplicity of 5, the jet $p_{\rm T}$ and rapidity $|$y$|$ for the 3 leading jets, and the jet H$_{\rm T}$ for a multiplicity up to at least 3 jets. The data distributions are corrected for all detector effects by means of regularised unfolding and compared with the particle level predictions by \mbox{\textsc{mg5\_amc@nlo}}~\cite{amcatnlo} and by \mbox{\textsc{mg5\_amc}} tree level at LO accuracy. The measured data is compared with a calculation at NNLO accuracy for W+1-jet production. The predictions describe data well on the exclusive and inclusive jet multiplicities within the uncertainties, and are in good agreement with data on the jet $p_{\rm T}$ spectra. The jet $|$y$|$ and H$_{\rm T}$ spectra are well modeled by both \mbox{\textsc{mg5\_amc@nlo}} prediction for all inclusive jet multiplicities and NNLO calculation for one inclusive jet. Overall, \mbox{\textsc{mg5\_amc}} tree level slightly underestimates data on the observables.

Four-lepton production in proton-proton collisions, pp $\rightarrow$ Z/$\gamma^*$ Z/$\gamma^*$ $\rightarrow$ $\ell\ell\ell\ell$ (where $\ell$ = e or $\mu$) is also studied~\cite{smp16001} with a data sample corresponding to an integrated luminosity of 2.6 fb$^{-1}$. The ZZ production cross section, measured for both Z bosons produced in the mass region 60 $<$ m$_Z<$ 120 GeV, is 14.6$^{+1.9}_{-1.8}$(stat)$^{+0.5}_{-0.3}$(syst)$\pm$0.2(theo)$\pm$ 0.4(lumi) pb. 
The results are consistent with SM predictions.

The WZ production cross section has been measured using a data sample corresponding to an integrated luminosity of 2.3 fb$^{-1}$~\cite{smp16002}. The measurement is performed in the leptonic decay modes WZ $\rightarrow$ $\ell\nu\ell'\ell'$, where $\ell$,$\ell'$ = e,$\mu$. 
The measured total cross section is 39.9$\pm$3.2(stat)$^{+2.9}_{-3.1}$(syst)$\pm$0.4(theo)$\pm$1.3 pb for the dilepton mass range 60 $<$ m$_{\ell\ell'}<$ 120 GeV. This measurement is consistent with the theoretical value of 
42.3$^{+1.4}_{-1.1}$(scale)$\pm$0.6(PDF) pb 
calculated at NLO, and with the NNLO prediction. 

The increased center of mass energy of the LHC Run II allows for a much larger reach to study possible deviations from the SM in a generic manner, using an effective field theory approach. In this analysis~\cite{smp16012} we constrain additional operators that would lead to anomalous WW$\gamma$ or WWZ couplings by studying events with one W boson decaying to an electron or $\mu$ and neutrino and one W or Z boson decaying hadronically. The study uses an integrated luminosity of 2.3 fb$^{-1}$. Using di-boson mass distributions we derive 95\% confidence intervals for the anomalous coupling parameters $\frac{c_{WWW}}{\Delta^2}$([-9.46,9.42] TeV$^{-2}$), $\frac{c_W}{\Delta^2}$ ([-12.6,12.0] TeV$^{-2}$) and $\frac{c_B}{\Delta^2}$ (-56.1,55.4] TeV$^{-2}$), in agreement with SM expectations. While the current limits cannot supersede the results from the LHC Run I [3], it is expected that limits will substantially improve with accumulating luminosity.

\section{Higgs Boson}
\begin{figure}[htb]
\centering
\includegraphics[height=2.2in]{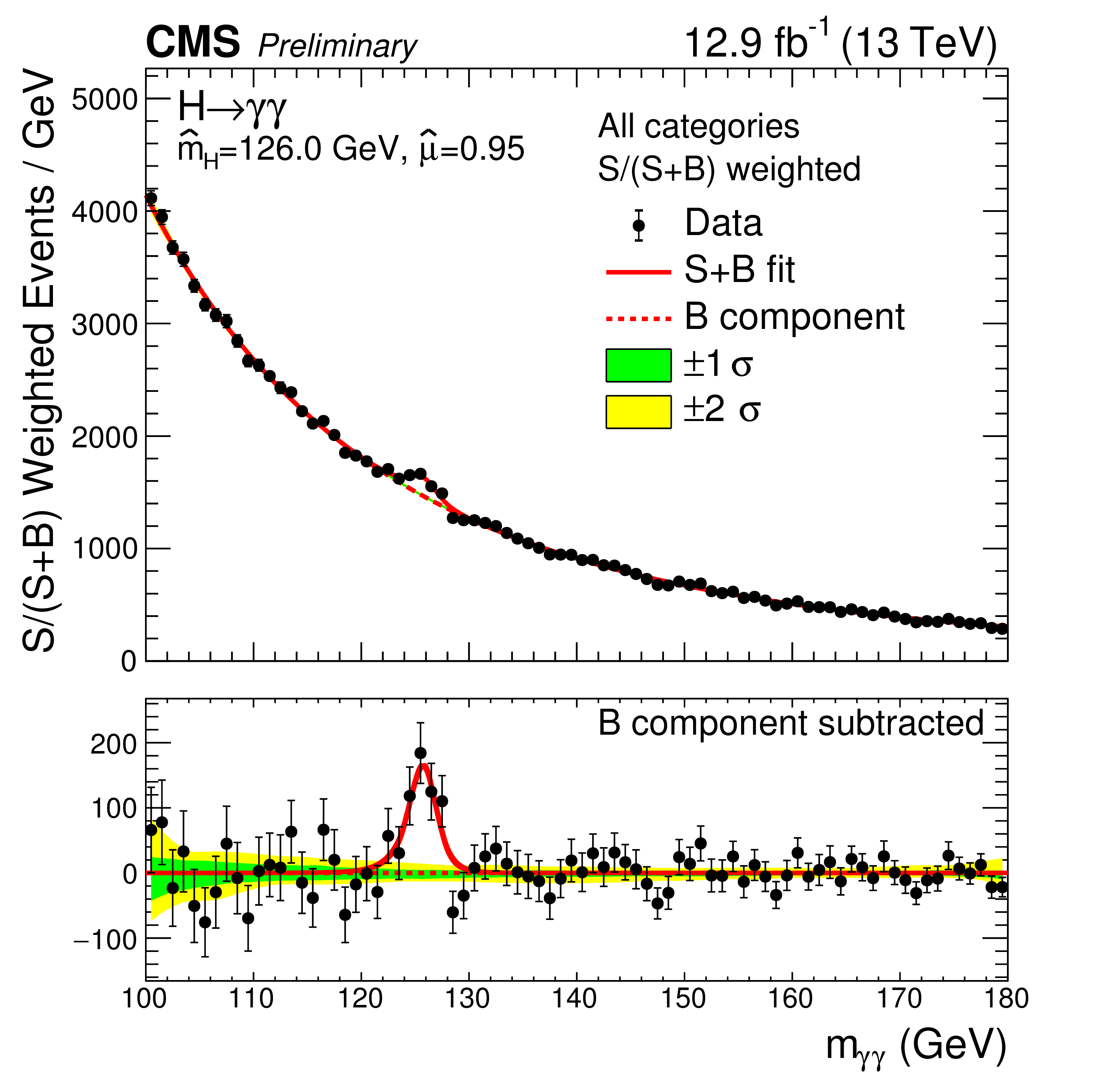}
\includegraphics[height=2.2in]{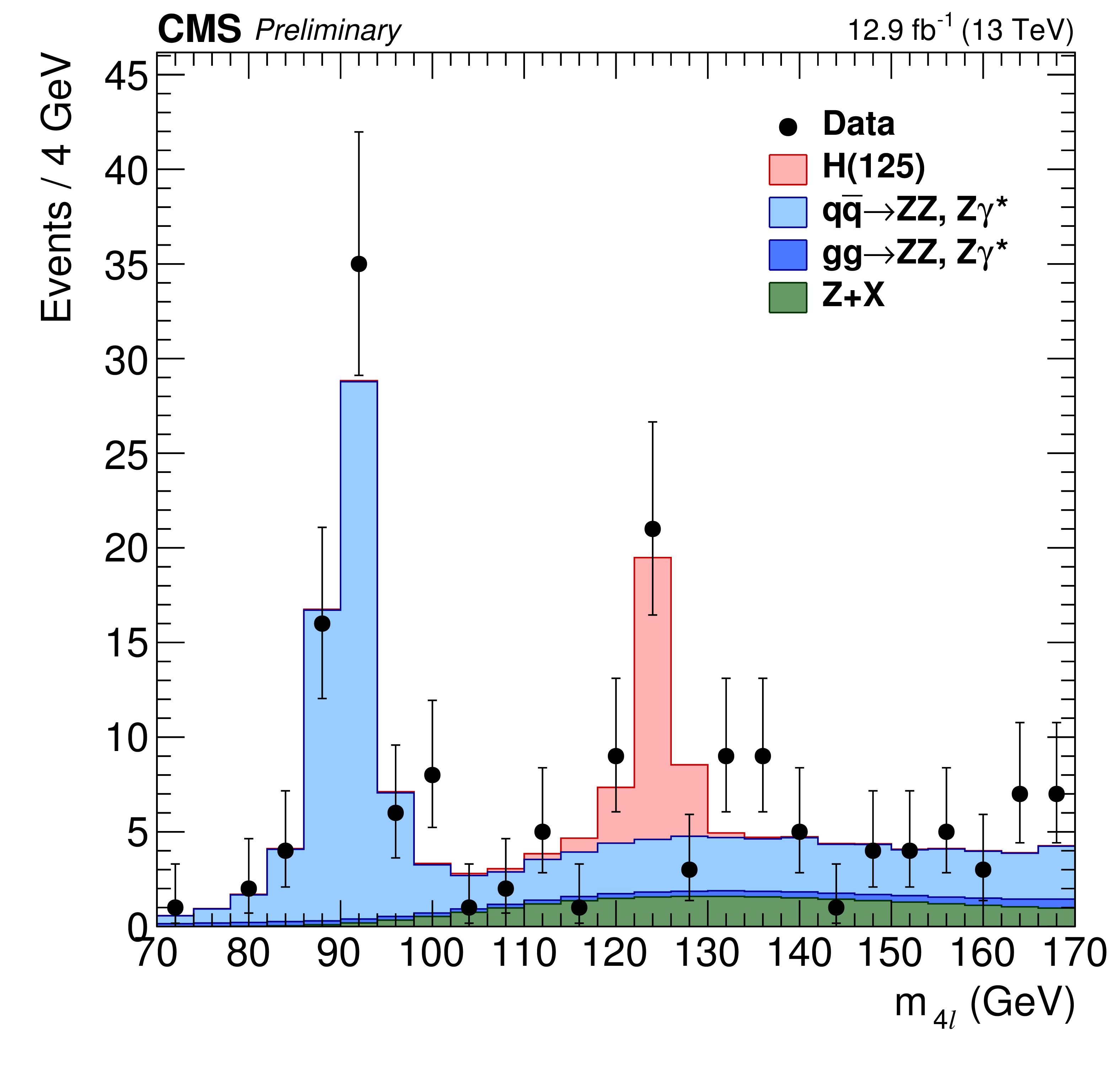}
\caption{Left (right): the Higgs boson signal in the di-photon~\cite{hig16020} (four-lepton~\cite{hig16033}) decay channel, obtained with 12.9 fb$^{-1}$ of 13 TeV data.}
\label{fig:higgs}
\end{figure}

Higgs boson production for the 2 photon decay~\cite{hig16020} and four leptons decay~\cite{hig16033} channels with 12.9 fb$^{-1}$ of the 2016 LHC Run 2 data has been observed, see Fig.~\ref{fig:higgs}. 

The observed significance in the 2 photon decay channel for the SM Higgs boson at the Run 1 ATLAS+CMS combined m$_H$=125.09 GeV is 5.6$\sigma$, where 6.2$\sigma$ is expected. A maximum significance of 6.1$\sigma$ is observed at 126.0 GeV. The best-fit signal strength relative to the SM prediction is 0.95$\pm$0.17(stat)$^{+0.10}_{-0.07}$(syst)$^{+0.08}_{-0.05}$(theo) when the mass parameter is profiled in the fit, and 0.91$\pm$0.17(stat)$^{+0.09}_{-0.07}$(syst)$^{+0.08}_{-0.05}$(theo) when it is fixed to m$_H$=125.09 GeV. The fiducial cross section is measured to be 69$^{+16}_{-22}$(stat)$^{+8}_{-6}$(syst) fb, where the SM theoretical prediction is 73.8$\pm$3.8 fb.

Higgs boson properties are studied using the H $\rightarrow$ ZZ $\rightarrow$ 4$\ell$ ($\ell$=e,$\nu$) decay channel. The observed significance for the SM Higgs boson with m$_H$ = 125.09 GeV is 6.2$\sigma$, where the expected significance is 6.5$\sigma$. The signal strength modifier $\mu$, defined as the production cross section of the Higgs boson times its branching fraction to four leptons relative to the SM expectation, is measured to be $\mu$=0.99$^{+0.33}_{-0.26}$ at m$_H$=125.09 GeV. The signal-strength modifiers for the main Higgs boson production modes have also been constrained. The model independent fiducial cross section is measured to be 2.29$^{+0.74}_{-0.64}$(stat.)$^{+0.30}_{-0.23}$(sys.)$^{+0.01}_{-0.05}$(model dep.) fb and differential cross sections as a function of the $p_{\rm T}$ of the Higgs boson and the number of associated jets are determined. The mass is measured to be m$_H$ = 124.50$^{+0.48}_{-0.46}$ GeV and the width is constrained to be $\Gamma_H<$41 MeV. The anomalous effects in the Higgs interactions with a pair of Z bosons are constrained under the assumption of a spin-zero resonance. Finally, a search for an additional resonance decaying to ZZ is performed for a range of masses up to 2.5 TeV and with various widths and no significant excess is observed.

\section{Exotica}

\begin{figure}[htb]
\centering
\includegraphics[height=2.2in]{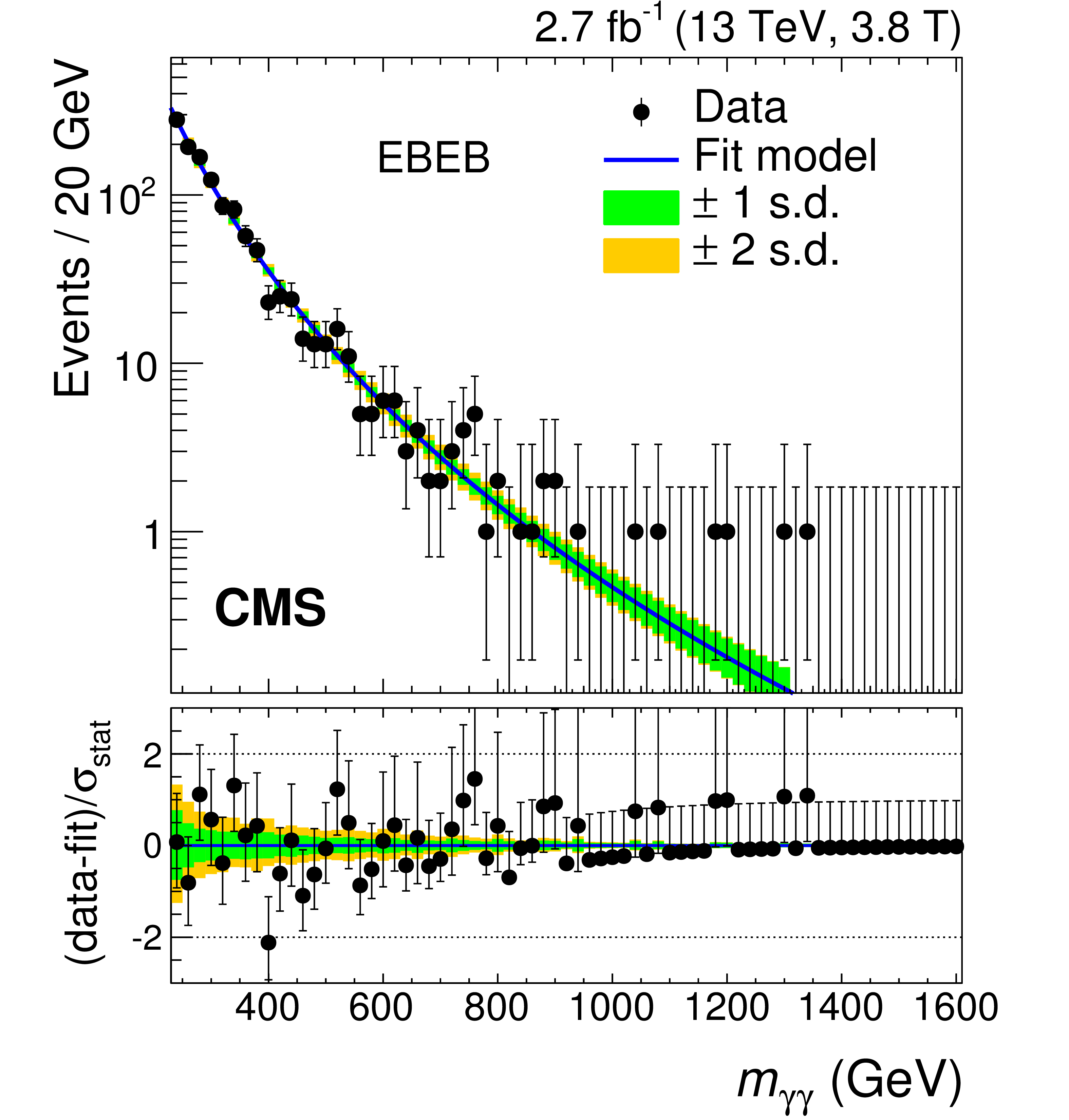}
\includegraphics[height=2.2in]{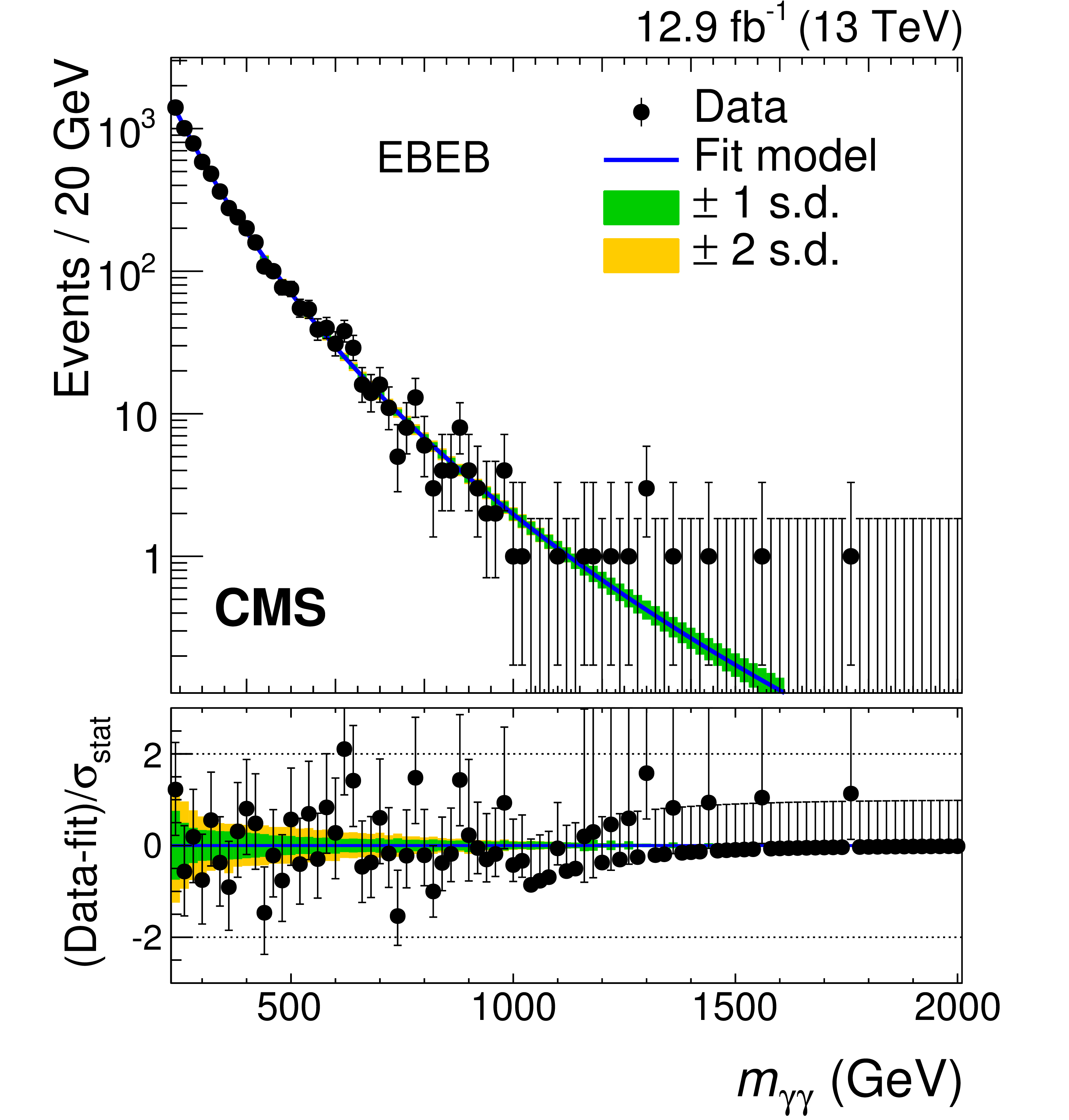}
\caption{Left (right): di-photon mass spectrum obtained with 2.7 fb$^{-1}$ (12.9 fb$^{-1}$) of 2015~\cite{exo2015} (2016~\cite{exo16027}) data at 13 TeV (both photons in the ECAL barrel).}
\label{fig:diph}
\end{figure}

The search for resonant production of high mass photon pairs with the 2015 dataset~\cite{exo2015} presented a hint of deviation from SM, see Fig.~\ref{fig:diph} (left). The same search has been performed using 12.9 fb$^{-1}$ of data~\cite{exo16027}. It is aimed at spin-0 and spin-2 resonances of mass between 0.5 and 4.5 TeV and width, relative to the mass, up to 5.6×10$^{-2}$. The results of the search are combined statistically with those previously obtained by the CMS collaboration at $\sqrt{s}$ = 8 and 13 TeV. The mild excess near m$_{\gamma\gamma}\sim$ 750~GeV reported by CMS with 2012 and 2015 data is not confirmed with 2016 data. No significant excess is observed over the SM predictions. Limits are set on scalar resonances produced through gluon-gluon fusion, and on Randall-Sundrum gravitons. 

Two searches for narrow resonances decaying to dijet final states with 12.9 fb$^{-1}$ of data are pursued~\cite{exo16032}. A low-mass search, for a resonance mass between 0.6 TeV and 1.6 TeV, is performed using dijets that are reconstructed from calorimeter information in the high-level trigger. A high-mass search, for resonances with mass above 1.6 TeV, is performed using dijets reconstructed with the particle flow algorithm from the normal reconstruction chain. 
The spectra are well described by a smooth parameterization and no significant evidence for new particle production is observed. Upper limits at 95\% CL are reported on the production cross section times branching ratio to dijets times acceptance of the $|\Delta_{jj}|$ and $|\eta|$ cuts for narrow resonances from quark-quark, quark-gluon and gluon-gluon final states. When interpreted in the context of specific models, the limits exclude string resonances with masses below 7.4 TeV, scalar diquarks below 6.9 TeV, axigluons and colorons below 5.5 TeV, excited quarks below 5.4 TeV, color-octet scalars below 3.0 TeV, W' bosons below 2.7 TeV, Z' bosons below 2.1 TeV and between 2.3 to 2.6 TeV, and RS gravitons below 1.9 TeV, extending previous  limits. 

A search for new narrow resonances in dielectron and dimuon spectra is performed using an integrated luminosity of 12.4 fb$^{-1}$ and 13.0 fb$^{-1}$, respectively~\cite{exo16031}. The observations are consistent with the expectations of the SM. Upper limits at 95\% CL are set on the cross section times branching fraction for new boson production relative to SM Z boson production. For the Z'$_{SSM}$ and Z'$_{\Psi}$ bosons, we obtain lower mass limits of 4.0 TeV and 3.5 TeV, respectively, which extend the limits from previous searches. 

The dark matter (DM) searches~\cite{CMS-DP-2016-057} use benchmark simplified models whith a minimal set of parameters including coupling structure, the mediator and DM masses $M_{\rm Med}$ and $m_{\rm DM}$, and coupling of mediator to SM and DM particles $g_{\rm q}$ and $g_{\rm DM}$. Di-jet searches and mono-X searches are interpreted using a DM model with a leptophobic axial vector mediator, assuming g$_q$ = 0.25 and g$_{\rm DM}$ = 1; the dijet searches are complimentary to the mono-X searches and cover the off-shell region that mono-X is less sensitive to. Mono-X searches assuming $g_{\rm q}$ = 1 and $g_{\rm DM}$ = 1 are also pursued; for smaller $M_{\rm Med}$, the mono-t$\rm\bar t$ search already has better sensitivity than the mono-jet search even with the 2015 dataset. Searches for DM with various mono-X final states have been performedand the data is found to be in agreement with the SM prediction. We expect updates with the full 2016 dataset in the near future.

\section{SUSY}

The substantial increase in the available dataset allowed a significant extension in the sensitivity for searches for supersymmetry. These searches have been designed and optimized for a wide range of SUSY models. In all analyses, the observed yields in the signal regions are found to be compatible with expectations from SM processes.

For R-parity conserving SUSY models, and with the $\tilde\chi_1^0$ as LSP, models of gluino and squark pair production are investigated. Within the context of the simplified models used for interpretation, mass limits at 95\% CL reach as high as 1800~\cite{sus16030}, 900~\cite{sus16015}, and 1400 GeV for gluinos, top squarks, and mass-degenerate first or second generation squarks, respectively. Scenarios with compressed mass spectra, with mass differences between the top squark and the LSP below $M_{\rm t}$ or $M_{\rm W}$, have also been searched for, either in decay chains of other sparticles, or by focussing on events with hard initial state radiation. Here the mass limits for the lightest top squark reach up to about 450 GeV~\cite{sus16029}. Electroweak production modes have been studied for the first time in Run II. Limits are set for the production of $\tilde\chi_1^{\pm}$$\tilde\chi_2^0$ pairs under different assumptions for their decay modes. Depending on these assumptions, 95\% CL mass limits reach up to approximately 1 TeV~\cite{sus16024}. Compressed mass spectra have also been investigated for this production mode, again using the presence of jets from initial state radiation and extending the acceptance to low lepton transverse momenta. 

The data expected in the remaining part of 2016 and in the following years will allow an extension of the sensitivity of existing searches and a broadening of the SUSY search programme in the near future.


\end{document}